\documentclass[sn-mathphys]{sn-jnl}
\usepackage{enumerate}
\usepackage{amsmath}
\usepackage{amsfonts}
\usepackage{amssymb}
\usepackage[utf8]{inputenc}
\usepackage[T1]{fontenc}
\usepackage{mathtools}
\usepackage{wasysym}
\usepackage{accents}
\usepackage{natbib}
\usepackage{subfig}
\usepackage{bm}

\jyear{2021}%

\theoremstyle{thmstyleone}%
\theoremstyle{thmstyletwo}%

\theoremstyle{thmstylethree}%

\begin{document}

\title[Central charge criticality of charged AdS black hole surrounded by different fluids]{Central charge criticality of charged AdS black hole surrounded by different fluids}


\author[1]{\fnm{R.} \sur{B. Alfaia}}\email{alfaiaramon@gmail.com}
\equalcont{These authors contributed equally to this work.}

\author*[2]{\fnm{I.} \sur{P. Lobo}}\email{lobofisica@gmail.com}
\equalcont{These authors contributed equally to this work.}

\author[1]{\fnm{L.} \sur{C. T. Brito}}\email{lcbrito@ufla.br}
\equalcont{These authors contributed equally to this work.}

\affil[1]{\orgdiv{Physics Department}, \orgname{Federal University of Lavras}, \orgaddress{\street{Avenida Norte}, \city{Lavras}, \postcode{37200-000}, \state{MG}, \country{Brazil}}}

\affil*[2]{\orgdiv{Department of Chemistry and Physics}, \orgname{Federal University of Para\'iba}, \orgaddress{\street{Rodovia BR 079 - Km 12}, \city{Areia}, \postcode{58397-000}, \state{PB}, \country{Brazil}}}


\abstract{We analyze the extended phase space thermodynamics of Kiselev black hole introducing a central charge and allowing the gravitational constant to vary. We also discuss the relation between the chemical potential and the size of the black hole, besides the new description of phase transitions. We obtain as a conclusion that the universality of the central charge does not remain valid in general.}




\maketitle

\section{Introduction}
Since the fundamental works of Bekenstein \cite{Bekenstein:1972tm,Bekenstein:1973ur}, Bardeen, Carter and Hawking \cite{Bardeen:1973gs,Hawking:1975vcx}, black hole thermodynamics have played a prominent role in theoretical physics due to its property of bringing together gravity, quantum physics and thermodynamics into a unified framework. Since then, many developments have been achieved as the existence of phase transitions when different kinds of black hole elements are considered, like the presence of a cosmological constant \cite{Hawking:1982dh}, electric charge, and many other types of matter Lagrangians \cite{Cai:2004eh,delaCruz-Dombriz:2009pzc,Faraoni:2010yi,MoraisGraca:2017nlv,Myers:1988ze,Blas:2011ni,Wald:1999vt}.
\par
In particular, one of the most interesting aspects that recently emerged consists in the possibility of assuming a negative cosmological constant $\Lambda$ as a thermodynamic variable, which plays the role of a pressure term. In fact, the use of the cosmological constant as dynamical variable stems from the original works of Teitelboim and Brown \cite{Teitelboim:1985dp,Brown:1988kg}, in which a quantum process of radiation of bubbles is the physical process responsible for spontaneously reducing $\Lambda$, which is an integration constant, or a ``hair'', in the metric solution discussed. The recognition of this quantity as a thermodynamic variable that should modify the first law was proposed years later in \cite{Creighton:1995au}. In this case, the fact that $\Lambda$ is an integration constant rather than a fundamental field allows one to consider geometries in which it assumes different values. In this context,  we now identify the mass parameter as an enthalpy instead of the internal energy, since the conjugate variable to $\Lambda$ is the volume of the Euclidean sphere \cite{Kastor:2009wy,Kubiznak:2012wp} (at constant pressure, these quantities coincide). In this case, intriguing features have been analyzed like the resemblance of the Reisser-Nordstr\"{o}m-anti de Sitter spacetime with a van der Waals fluid as a thermodynamic system, a property that also allowed this approach to be named \textit{black hole chemistry} \cite{Kubiznak:2016qmn}. We can see the productiveness of this research area by the features that have been extensively analyzed in the literature in the past years \cite{Fernando:2016sps,Ma:2017pap,Hendi:2015kza,Majhi:2016txt,Lobo:2017dib} as, for example,  the existence and properties of phase transitions, classification of phases, conjectured geometric inequalities and the effect of different kinds of matter fields and quantum effects as conductors of transitions.
\par
Due to the appearance of a negative cosmological constant, i.e., the fact that we are dealing with a black hole in anti-de Sitter (AdS) spacetime, the relation between this approach and the holographic principle naturally emerges. Based on this, the system's behavior as a holographic heat engine \cite{Johnson:2014yja} has also been studied in the literature considering the effect of different contributions on the efficiency of some cycles \cite{EslamPanah:2020hoj,Singh:2020xju,MoraisGraca:2018ofn,Bezerra:2019qkx,Johnson:2015fva}. However, a satisfactory realization of this principle in black hole thermodynamics has been a matter of scrutiny \cite{Karch:2015rpa}. Recently, advances in this direction have considered the possibility of considering simultaneously the cosmological and gravitational constants, $\Lambda$ and $G$, as thermodynamic variables. This approach relies on the introduction of a new thermodynamic quantity: the central charge $C$ of the dual Conformal Field Theory (CFT).  The central charge measures the degrees of freedom of the corresponding CFT in the AdS/CFT correspondence approach, which is related to the number of colors in the dual gauge theory. For this reason, its conjugate potential is referred as a chemical potential, which, in a usual thermodynamic system, is conjugate to the number of particles \cite{Kastor:2014dra}. So, in order to move from one CFT to another, the central charge should be allowed to vary. However, this quantity is a function of the cosmological and gravitational constants. On the other hand, a variation of the cosmological constant (AdS radius) is related to a variation of the CFT volume \cite{Cong:2021fnf}, which is an independent thermodynamic quantity itself. Therefore, in order to vary, in an independent way, the cosmological constant and the central charge, it becomes necessary to vary the gravitational constant: the route towards a unified description of black hole thermodynamics in AdS space and the corresponding CFT thermodynamics requires the variation of $G$ as has been argued in some recent papers, see for instance \cite{Visser:2021eqk,Cong:2021jgb} and references therein.
\par
Also, from a classical gravitational perspective, the Jordan frame of scalar-tensor theories can be interpreted as the realization of the idea of a spacetime-varying gravitational constant due to the nonminimal coupling between the scalar field (gravitational constant) and the curvature scalar \cite{Faraoni:2004pi}. Motivations for considering this scenario stem from theoretical grounds, in which this approach would be closer to Machian ideas than General Relativity \cite{Brans:1961sx}, but also from experimental grounds, in which possible variations of the gravitational constant are being searched by different methods (see \cite{Uzan:2010pm} for a review on this subject). For instance, recently, a novel approach has been addressed and suggests a $5\%$ variation of $G$ in different environments \cite{Desmond:2020nde}, namely the Solar System and the Large Magellanic Cloud (outside of the Milky Way).
\par
Other reasons for investigating consequences of the noncontancy of the gravitational constant stems from a field theory perspective based Renormalization Group equations. From the effective quantum field theory related to the Einstein-Hilbert action \cite{Donoghue:1994dn} we learn that although the running of $G$ is inconclusive perturbatively \cite{Anber:2011ut}, it may have interesting properties in the non-perturbative regime \cite{Reuter:2001ag}.
\par
Based on this approach, the asymptotic safety program describes, in effective way, quantum gravitational effects by running the coupling constants of the classical gravity theory, namely, the gravitational constant $G\rightarrow G(k)$ and cosmological constant $\Lambda\rightarrow \Lambda(k)$, where $k$ is the Renormalization Group scale \cite{Niedermaier:2006ns,Adeifeoba:2018ydh} (for phenomenological opportunities of this approach to quantum gravity, we refer the reader to \cite{Eichhorn:2018yfc,Addazi:2021xuf}).
\par
The analysis presented in \cite{Cong:2021fnf} contains interesting results that may open a new branch in black hole thermodynamics. From a fundamental point of view, some aspects which deserve attention are the conjectured role played by the central charge in phase transitions and the existence of a ``universal'' critical charge from which phase transitions can occur. Besides that, there are other interesting points to be highlighted: the realization of a holographic Smarr relation which leads to the first law of thermodynamics in the usual $D$-dimensional spacetime when $G$ is constant; the verification of the validity of the reverse isoperimetric inequality for a class of black holes; the redefinition of some quantities like a new thermodynamic volume and the introduction of a ``chemical potential''. In a few words, Ref.\cite{Cong:2021fnf} is based on the thermodynamics of Kerr-Newman-anti de Sitter black hole with a variable $G$ and the introduction of new thermodynamic variables that mix the cosmological constant, the spacetime dimension and the gravitational constant in a new extended phase space. In this work, we investigate if those results remain valid when a different matter content is present and what is the impact of this matter content on the findings of the original paper. To achieve this objective, we will use the so-called Kiselev black hole in $D$-dimensions \cite{Kiselev:2002dx,Chen:2008ra} as our ansatz, and for simplicity reasons, we will  assume the static configuration.
\par
Our paper is organized as follows. In section \ref{sec:kiselev}, we verify the validity of holographic formulation of the thermodynamics of Kiselev (Reissner-Nordstr\"{o}m-anti de Sitter) black hole. We consider a variable gravitational constant and the introduction of the central charge as a thermodynamic variable. In section \ref{sec:extended_phase}, we study the $4$-dimensional case in this new extended phase space structure and verify the relation between the size of the black hole and its chemical potential. In section \ref{sec:critical}, we study its critical behavior, the conditions for the existence of the universal nature of the critical central charge. Then, in section \ref{sec:conclusion}, we draw our final remarks. We assume natural units: $\hbar=c=k_{\text{B}}=1$.


\section{Kiselev black hole in arbitrary dimensions}\label{sec:kiselev}

To illustrate this extended thermodynamic approach, we are going to consider a solution of Einstein's field equations surrounded by matter content in which the gravitational and cosmological constants are allowed to vary, which could be realized, for instance in the Asymptotic Safety approach to quantum gravity by replacing the coupling constants of the theory by their corresponding running couplings $G(k)$ and $\Lambda(k)$, where $k$ describes the resolution scale at which the spacetime is probed \cite{Niedermaier:2006ns}. Therefore the action assumed is the Einstein-Hilbert one in $D$-dimensions with a matter Lagrangian minimally coupled to Riemannian geometry \cite{Eichhorn:2020mte}
\begin{equation}
    S=\int d^Dx\sqrt{-g}\left[\frac{1}{16\pi G}(R-2\Lambda)+{\cal L}_{\text{matter}}\right]\, ,
\end{equation}
where $G=G(k)$ is the $D$-dimensional gravitational constant and $\Lambda=\Lambda(k)$ is the running cosmological constant that are allowed to vary. From now on, we shall treat the scale $k$ as independent of the spacetime coordinates and we will omit the $k$-dependence in our notation. For thermodynamic purposes, we shall investigate the effects of the variation of these constants, as well as the usual black hole parameters. In fact, this implies that we can consider the usual solution of Einstein equations, but with $k$-dependent couplings \cite{Adeifeoba:2018ydh}.\footnote{In sections \ref{subsec:g_variable}, \ref{sec:extended_phase} and \ref{sec:critical}, we shall be considering the effects of assuming $G$ as a local thermodynamic variable that will contribute to the first law of thermodynamics, like the black hole area or electric charge.} 
\par
In this section, we consider the so-called Kiselev black hole \cite{Kiselev:2002dx}, that has been extensively analyzed in the literature that aims to incorporate the effects of exotic matter that, in a cosmological setting, would be responsible for describing an accelerated expansion of the universe,  which furnishes interesting phenomenological opportunities that have been recently investigated, for instance in \cite{Younas:2015sva,Rizwan:2018lht,Rayimbaev:2022mrk}. We assume the $D$-dimensional spherically symmetric spacetime ansatz
\begin{equation}
    ds^2=f(r)dt^2-g(r)dr^2-r^2d\theta_1^2-r^2\sin^2\theta_1 d\theta_2^2-...-r^2\sin^2\theta_1...\sin^2\theta_{D-3}d\theta^2_{D-2}\, ,
\end{equation}
and that the black hole is surrounded by matter with an average stress energy tensor of the form \cite{Chen:2008ra}
\begin{align}
    &T^{\ t}_{t}=T^{\ r}_r=\rho(r)\, ,\\
    &T_{\theta_1}^{\ \theta_1}=T_{\theta_2}^{\ \theta_2}=...=T_{\theta_{D-2}}^{\ \theta_{D-2}}=-\frac{\rho(r)}{D-2}[(D-1)\omega+1]\, ,
\end{align}
where $\rho$ is interpreted as an average energy density of this fluid and $\omega$ is an average equation of state parameter, that labels different fluids.\footnote{Let us be more explicit about the notation used: in fact, Kiselev's stress-energy tensor does not describe a perfect fluid, as highlighted in \cite{Visser:2019brz,Boonserm:2019phw}, however an average equation of state parameter can be defined, which is given by $\omega$. This issue is also discussed in the second section of \cite{Lobo:2020jfl}.} For instance, in four dimensions, the so called average quintessence matter ($-1<\omega<-1/3$) \cite{Sahni:2004ai} is described by $\omega=-2/3$, $\omega=-1$ describes an average cosmological constant and $\omega=-4/3$ describes an average phantom field ($\omega<-1$) \cite{Caldwell:1999ew}.
\par
The solution of these field equations obeys additivity and linearity properties. It means that we can add fluid contributions to the field equations by adding related extra terms to the metric. The solution of the Einstein's field equations that endows $n$ different matter contributions has been found in \cite{Kiselev:2002dx,Chen:2008ra} as
\begin{equation}
f(r)=g(r)^{-1}=1-\frac{r_g}{r^{d-3}}+\sum_{n}\frac{r_n}{r^{(D-1)\omega_n+D-3}}\, ,
\end{equation}
where $r_g$ is a function of the black hole's mass and $r_n$ are normalization constants. This solution can describe the usual Reissner-Nordstr\"{o}m-Tangherlini metric by assigning specific forms to $(r_g,r_n,\omega_n)$ \cite{Tangherlini:1963bw,Kunstatter:2002pj,Yang:2018cim,Wu:2017agu,Emparan:2008eg}
\begin{align}
    &f(r)=g(r)^{-1}=1-\frac{16\pi GM}{(D-2)\Omega_{D-2}r^{D-3}}+\frac{ 32\pi^2 (\sqrt{G}Q)^2}{(D-2)(D-3)\Omega_{D-2}^2  r^{2(D-3)}}\nonumber\\
   & -\frac{2\Lambda r^2}{(D-1)(D-2)}-\frac{G\,  b}{r^{(D-1)\omega+D-3}}\, ,\label{metric1}
\end{align}
where $G$ is the gravitational constant in arbitrary dimensions, $M$ is the black hole's mass, $Q$ is its electric charge, $\Lambda$ is the cosmological constant (that we shall assume as negative in this paper), $b$ is a parameter related to the energy density of Kiselev's fluid and $\Omega_{D-2}=2\pi^{(D-1)/2}{\Gamma((D-1)/2)}$ is the area of the unit sphere ${\mathbb S}^{D-2}$ (which in $4$-dimensions gives $4\pi$).
\par
We also have the form of each contribution of the energy density, in which we assumed the standard matter coupling in Einstein equations $G_{\mu\nu}=8\pi G\, T_{\mu\nu}$:
\begin{equation}
    \rho_n(r)=r_n\frac{\omega_n(D-1)(D-2)}{32\pi r^{(D-1)(\omega_n+1)}}\, .\label{rho1}
\end{equation}
Notice that the case of the equation of state parameter $\omega=0$ would be added to the Schwarzschild term in Eq.(\ref{metric1}). For the case $\omega=(D-3)/(D-1)$, the Kiselev term would mix with the Reissner-Nordst\"{o}m one. If $\omega=-1$, the dependence on the spacetime dimension $D$ would disappear in the Kiselev term of (\ref{metric1}) and it would contribute to the cosmological constant term.
\par
Notice, in this approach, that the cosmological constant was not introduced in the Einstein-Hilbert action as a fundamental piece, but instead, emerged as a special contribution of the Kiselev fluid.


\subsection{Thermodynamics when $G$ is a constant}

In this first subsection, we are going to express the usual thermodynamic quantities when $G$ is constant. It will be important to deduce the first law and Smarr formula for the $D$-dimensional Kiselev black hole as is done in \cite{Chen:2008ra}. Then, from the form of the first law with $G$ fixed, we shall verify what turns out to be when we let $G$ varies as a thermodynamic quantity. 
\par
As usual, the temperature reads
\begin{align}
    T=\frac{\kappa}{2\pi}=\frac{1}{4\pi}\left[\frac{df}{dr}\right]_{r=r_+}=\frac{1}{4\pi}\left[\frac{D-3}{r_+}-\frac{32\pi^2 G Q^2}{(D-2)\epsilon_0\Omega^2_{D-2} r_+^{2D-5}}\right.\nonumber\\
   \left. -2\frac{\Lambda}{D-2}r_+ +(D-1)\frac{G b\, \omega}{r_+^{D-2+\omega(D-1)}}\right]\, ,
\end{align}
where $\kappa$ is the surface gravity.
\par
The entropy is found by integrating the expression $dS=dM/T$ as
\begin{equation}
S=\frac{\Omega_{D-2}}{4G}r_+^{D-2}=\frac{A}{4G}\, ,
\end{equation}
where $A$ is the area of the sphere of radius $r_+$ in $D$ dimensions.
\par
The electric potential reads
\begin{equation}
    \phi=\frac{\partial M}{\partial Q}\biggr\rvert_{S,\Lambda,b}=\frac{4\pi Q}{(D-3)\Omega_{D-2}}r_+^{3-D}\, ,
\end{equation}
which gives exactly the electrostatic potential at $D=4$ in Gaussian units.
\par
The pressure is simply proportional to $\Lambda$ and volume is its conjugate quantity:
\begin{align}
   P=-\frac{\Lambda}{8\pi G}\doteq\frac{(D-1)(D-2)}{16\pi G\,  l^2}\, ,\label{pressure1}\\
   V=\frac{\partial M}{\partial P}\biggr\rvert_{S,Q,b}=\frac{\Omega_{D-2}}{D-1}r_+^{D-1}\, ,
\end{align}
where, in fact, $V$ is the Euclidean volume in $D$-dimensions, and we introduced the quantity $l$, that shall be convenient in the following sections (from \eqref{pressure1}, $l^2=-(D-1)(D-2)/(2\Lambda)$). Notice that this approach bears some similarities with proposals that considered a variation of the cosmological constant due to its introduction as a ``hair'', like the black hole's mass and charge \cite{Teitelboim:1985dp,Brown:1988kg}.
\par
The Kiselev (or quintessence) potential is
\begin{equation}
    B=\frac{\partial M}{\partial b}\biggr\rvert_{S,Q,P}=-\frac{(D-2)\Omega_{D-2}}{16\pi  }r_+^{\omega(1-D)}\, .
\end{equation}
\par
When $D=4$, these quantities reduce to the ones found in \cite{GBKiselev}. From \cite{Chen:2008ra}, we can see the Smarr formula and first law of thermodynamics takes the form
\begin{align}
    (D-3)M=(D-2)TS+(D-3)\phi Q-2PV +[\omega(D-1)+D-3]Bb\, ,\label{Smarr1}\\
    \delta M=T\delta S+V\delta P+\phi \delta Q+B\delta b\, .\label{usual1law}
\end{align}
\par
In principle, these equations do not depend on $G$, but if we intend to investigate the effects of allowing it to vary, we need to start from the first law with, indeed, the black hole quantities that appear in the metric function (\ref{metric1}). In fact, if we recover $G$, the first law assumes the form
\begin{equation}
    \delta(GM)=\frac{\kappa}{8\pi}\delta A-\frac{V}{8\pi}\delta \Lambda+\sqrt{G}\phi\delta(\sqrt{G}Q)+B\delta(G\, b)\, .\label{1law1}
\end{equation}
\par
This is the expression that is going to be analyzed in the following section.


\subsection{Assuming $G$ as a thermodynamic variable}\label{subsec:g_variable}

It has been argued \cite{Visser:2021eqk} that the Smarr relation that is supposed to be compatible with the holographic principle should be of the form $M=TS+\nu^i B_i+\mu C$, where $C$ is the central charge that relates to the cosmological constant and the gravitational constant as
\begin{equation}\label{centralcharge}
    C=k\frac{l^{D-2}}{16\pi G}\, ,
\end{equation}
where $k$ depends on the specifics of the holographic model \cite{Karch:2015rpa,Cong:2021fnf} and $\mu$ is its corresponding potential, that we shall interpret as a chemical potential using the language of the black hole chemistry. The terms $\nu^i$ correspond to additional potentials of the problem \cite{Visser:2021eqk}, in this case the electric and Kiselev ones. Notice that the term of the bulk pressure (cosmological constant) does not appear in this relation, not even the spacetime dimension is made explicit. In this paper, we shall not discuss the details or soundness of this proposal, but instead we will investigate the consequences of assuming it in the Kiselev background. Also notice that $C$ depends on the spacetime dimension $D$, the gravitational constant $G$ and also on the cosmological constant by means of the $l$ term (see Eq.(\ref{pressure1})).
\par
From Eq.(\ref{1law1}), if we let $G$ to vary, we find
\begin{equation}
     \delta M=\frac{\kappa}{8\pi G}\delta A-\frac{V}{8\pi G}\delta \Lambda+\phi\delta Q + B\delta b-\frac{\beta}{G}\delta G\, ,\label{1lawG}
\end{equation}
where 
\begin{equation}\label{beta1}
    \beta=M-\frac{1}{2}\phi Q-bB=TS+\frac{1}{2}\phi Q+\tilde{\mu}C\,.
\end{equation}
To obtain the last equation, we have used the proposed holographic Smarr relation
\begin{equation}\label{hSmarr}
M=TS+\phi Q+bB+\mu C\, .
\end{equation}
as a way to incorporate the central charge contribution. This is just a simple generalization of what was found in \cite{Cong:2021fnf}.
\par
In order to go deeper in the exploration of this formalism, let us find the length dimensions $L$ of the quantities involved in this framework (this can be seen from the thermodynamic quantities found in the previous section, from the metric function (\ref{metric1}) and from the assumption of natural units):
\begin{align}
&[V]=L^{D-1}\, , \ \ \   [S]=L^0\, , \ \ \   [A]=L^{D-2}=[G]\, , \ \ \ [\Lambda]=[\mu]^2=[M]^2=L^{-2}\, , \label{scaling1}\\
&[Q]=L^{(D-4)/2}\, , \ \ \ [\phi]=L^{(2-D)/2}\, , \ \ \ [b]=L^{\omega(D-1)-1}\, , \ \ \ [B]=L^{\omega(1-D)}\, .\nonumber
\end{align}

From (\ref{1lawG}), (\ref{beta1}) and (\ref{scaling1}), we derive the Smarr relation (now considering $G$ and $\Lambda$) for variable $G$ as
\begin{equation}
    -M=\frac{V}{4\pi G}\Lambda-\phi Q-(D-2)\mu C+[\omega(D-1)-1]bB\, ,
\end{equation}
from which we find
\begin{equation}
    V=\frac{8\pi G l^2}{(D-1)(D-2)}\{M-\phi Q-(D-2)\mu C+[\omega(D-1)-1]bB\}\, .
\end{equation}
\par
Again, using the holographic Smarr relation (\ref{hSmarr}) (in which the information about $D$, $G$ and $\Lambda$ is inserted in the thermodynamic quantity $C$), this can be simplified as
\begin{align}
    V=\frac{1}{2P}\{(D-2)TS-(D-3)(TS+\mu C+bB)+[\omega(D-1)+D-3]bB\}\nonumber \\
    =\frac{1}{2P}\{(D-2)TS-(D-3)\phi Q+[\omega(D-1)+D-3]bB-(D-3)M\}\, ,
\end{align}
where we used again (\ref{hSmarr}). This is exactly the Smarr relation (\ref{Smarr1}) when $G$ is a constant. Therefore, we demonstrate the preservation, for different fluids, of the result found in \cite{Cong:2021fnf} about the compatibility between the {\it holographic Smarr relation}, the {\it first law of black hole thermodynamics in $D$-dimensions with variable $G$} and the {\it Smarr relation in $D$-dimensions for constant $G$}.
\par
In order to properly write down this approach using the pressure and the central charge, we see that from Eq.(\ref{centralcharge}) we have
\begin{equation}
    \frac{\delta G}{G}=-\frac{2}{D}\frac{\delta C}{C} -\frac{(D-2)}{D}\frac{\delta P}{P}\, ,
\end{equation}
which allows us to write the bulk first law (\ref{1lawG}) as
\begin{equation}
    \delta M=T\delta S+\phi \delta Q+ B\delta b +V_C\delta P + \mu \delta C\, ,
\end{equation}
where
\begin{equation}\label{new_variables1}
    V_C=\frac{2M+(D-4)\phi Q+2[\omega(D-1)-1]bB}{2DP}, \ \ \ \ \mu=\frac{2P(V_C-V)}{C(D-2)}\, .
\end{equation}

The term $V_C$ is the new thermodynamic volume, that generalizes the one found in \cite{Cong:2021fnf}, and $\mu$ is the conjugate quantity that furnishes the response of the system to variations in the central charge. As can be verified, if $\delta G=0$, one recovers the usual first law of thermodynamics (\ref{usual1law}), since $\delta C$ and $\delta P$ must be related in this case.


\section{Extended phase space thermodynamics with variable $G$}\label{sec:extended_phase}

In order to apply the techniques explored in the previous section, we resort to the $4$-dimensional case. So, consider the metric function (\ref{metric1}) when $D=4$ and the relation between $\Lambda$ and $l$ from (\ref{pressure1}):
\begin{equation}\label{metric2}
    f(r)=1-\frac{2GM}{r}+\frac{GQ^2}{r^2}+\frac{r^2}{l^2}-\frac{Gb}{r^{3\omega+1}}\, .
\end{equation}
\par
The standard thermodynamic quantities can be straightforwardly calculated:
\begin{align}
    &M=\frac{r_+}{2G}+\frac{Q^2}{2r_+}+\frac{r_+^3}{2Gl^2}-\frac{b}{2r_+^{3\omega}}\, , \ \ \ T=\frac{3r_+^4+l^2r_+^2-GQ^2l^2+3Gbl^2\omega r_+^{1-3\omega}}{4\pi l^2r_+^3}\, , \label{4d1}\\ 
    &S=\frac{\pi r_+^2}{G}\, , \ \ \ V=\frac{4\pi}{3}r_+^3\, , \ \ \ \phi=\frac{Q}{r_+}\, , \ \ \ B=-\frac{1}{2r_+^{3\omega}}\, .
\end{align}
\par
As can be seen, since in $4$ dimensions, $G$ has dimensions of area, the quantities above have length dimensions $L$ compatible with the relations \eqref{scaling1} when $D=4$ (we also have $[T]=L^{-1}$). The new variables defined in (\ref{new_variables1}):
\begin{align}
    V_C=\frac{\pi}{3r_+}(GQ^2l^2+l^2r_+^2+r_+^4-3Gbl^2\omega r_+^{-3\omega+1})\, , \label{new_variables2-1}\\
    \mu=\frac{2\pi}{k l^4r_+}(GQ^2l^2+l^2r_+^2-3r_+^4-3Gbl^2\omega r_+^{-3\omega+1})\, .\label{new_variables2-2}
\end{align}

From the above expressions, we see that if $b\,\omega<0$, the thermodynamics volume $V_C$ is a positive definite quantity, and this condition coincides with the one that requires the positiveness of the energy density of Kiselev fluid (\ref{rho1}). Since $P$ and $C$ are positive quantities, we see that in the phase space region in which $V_C>V$, the following reverse isoperimetric inequality  \cite{Cong:2021fnf,Kubiznak:2016qmn,Cvetic:2010jb}
\begin{equation}
    \left[\frac{(D-1)V}{\Omega_{D-2}}\right]^{\frac{1}{D-1}}\geq \left(\frac{A}{\Omega_{D-2}}\right)^{\frac{1}{D-2}}\label{ineq1}
\end{equation}
will remain valid for the new thermodynamic volume $V_C$. As shown in \cite{Azreg-Ainou:2014lua}, for the usual Kiselev black hole volume, this inequality is valid for $ 1\geq (-\omega)^{1/3}$. 
\par
Now, let us investigate the sign of the chemical potential $\mu$ and its relation to the black hole size and the reverse isoperimetric inequality for some specific cases of Kiselev fluids. We are going to analyze cases of exotic matter components that surround the black hole, which effectively describe different rates of expansion if applied to a cosmological scenario. The cosmological quintessence matter \cite{Sahni:2004ai} is usually described by an equation of state parameter $\omega\in (-1,-1/3)$ with the objective of describing the accelerated expansion of the universe. This means that our case $\omega=-2/3$ aims to mimic this case of a quintessential fluid, however surrounding the black hole.
\par
The case $\omega=-1/3$ consists of the limiting case of the above inequality, while $\omega=-1$ (cosmological constant) describes the other bound. We do not analyze the case $\omega=-1$ individually, because it could be described by just shifting the value of the cosmological constant.
\par
\begin{figure}
    \centering
    \subfloat[][\centering We set $Q=G\ell_0^{-2}=1$, $l/\ell_0=7.8$, $b\ell_0^{1-3\omega}=10^{-3}$ and the central charge parameter $k=16\pi$.]
{\includegraphics[scale=0.9]{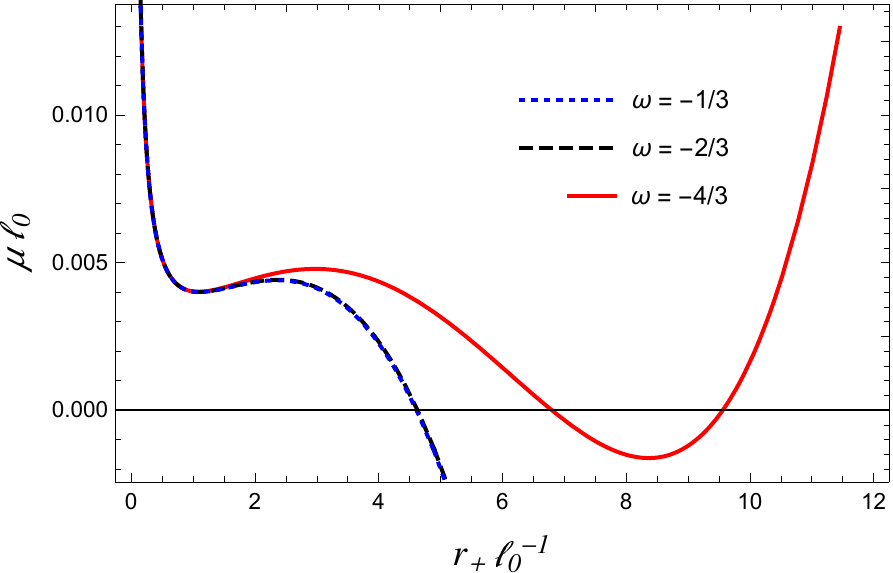}\label{fig:mu123_1}}
\hfill%
    \subfloat[][\centering We set $Q=0.05$, $G\ell_0^{-2}=1$, $l/\ell_0=2$, $b\ell_0^{1-3\omega}=10^{-1}$ and the central charge parameter $k=16\pi$. The red curve grows indefinitely.]{\includegraphics[scale=0.9]{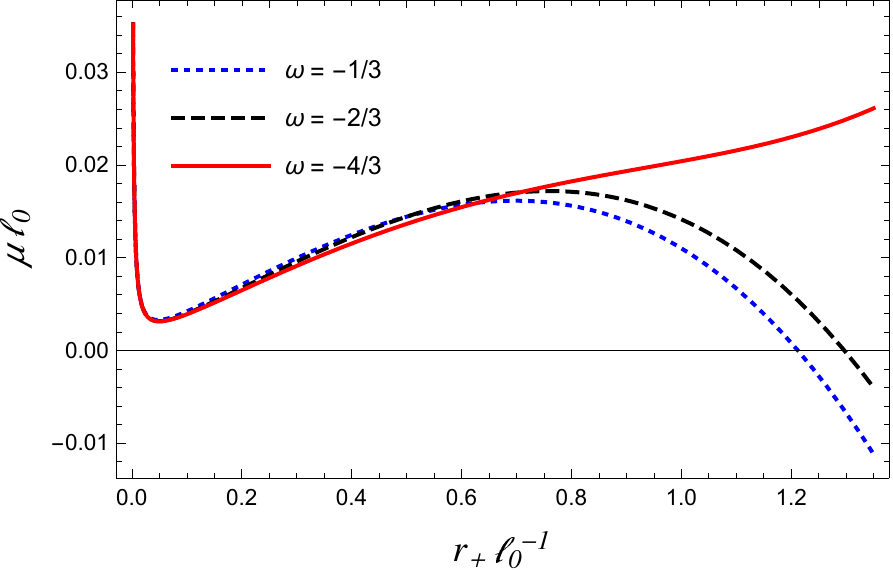} \label{fig:mu123_2}}
    \caption{Depiction of $\mu(r_+)\, \times\,  r_+$ in units of a fiducial length $\ell_0$.}
    \label{fig:mu123}
\end{figure}

Besides that, a super-negative equation of state parameter is described by $\omega < -1$ and is called phantom field and has also been considered in expanding cosmological scenarios \cite{Caldwell:1999ew}. In our paper, this case is represented by the $\omega=-4/3$.
\begin{itemize}
    \item \bm{$\omega=-1/3$}. In this case, the standard isoperimetric inequality is satisfied and it is possible to directly relate the size of the black hole and the sign of the chemical potential. In fact, one has $\mu>0$ for small black holes 
    \begin{equation}
        r_+^2<\frac{l^2}{6}\left[(1+Gb)+\sqrt{1+Gb+12\frac{GQ^2}{l^2}}\right]\, ,
    \end{equation}
which also satisfies the inequality (\ref{ineq1}) for the new thermodynamic volume $V_C$, since in this case $V_C>V$. On the other hand, large black holes are characterized by $\mu<0$. This classification and its relation with the isoperimetric inequality is compatible with the results found in \cite{Cong:2021fnf}. The behavior of $\mu(r_+)$ can be seen in Fig.(\ref{fig:mu123}) in the blue/dotted line, where we described the dimensioful functions in terms of a fiducial length $\ell_0$. 

\item \textbf{Average quintessence} \bm{$\omega=-2/3$}. In this case, the inequality involving $r_+$ that separates positive and negative values of $\mu$ is quite complicated. Nevertheless, we can see that the term $-3r_+^4$ in (\ref{new_variables2-2}) will dominate for very large black holes. Therefore, we verify the same classification above, i.e., that small black holes present $\mu>0$, while large ones present $\mu<0$. Also, the reverse isoperimetric inequality is satisfied for $V$ and will be valid for $V_C$ for small black holes. The behavior of $\mu(r_+)$ can be seen in Fig.(\ref{fig:mu123}) in the black/dashed line.
\par
As a matter of fact, this behavior will occur whenever we have $-1<\omega\leq 0$. If $\omega=-1$, this behavior will also occur as long as $Gb<l^{-2}$, in order to preserve the anti-de Sitter-like metric contribution (\ref{metric2}).

\item \textbf{Average phantom field}  \bm{$\omega=-4/3$}. In this case, one has $\mu>0$ for
\begin{equation}
    m(r_+)=GQ^2l^2+l^2r_+^2-3r_+^4+4Gbl^2 r_+^{5}>0\, ,
\end{equation}
which is achievable in a more diverse regime than the previous case. One can have $\mu>0$ for small and very large black holes, depending on the parameters considered. Therefore, in this case, the sign of $\mu$ does not work for this classification of black hole sizes, as can be seen in Fig.(\ref{fig:mu123}) in the red/solid curve. Curiously in this case, the reverse isoperimetric inequality (\ref{ineq1}) is no longer valid for the standard volume $V$.
\end{itemize}

\section{Critical behavior}\label{sec:critical}
As also happened in the original formulation of black hole chemistry, we can also study phase transitions induced by the presence of the black hole  electric charge. The critical quantities are defined by the saddle points of our thermodynamic functions. In this case, it will be convenient to study the equation of state $T=T(r_+)$ found in (\ref{4d1}) and to solve the following system of equations
\begin{equation}
    \frac{\partial T}{\partial r_+}\biggr\rvert_{r_+=r_c,l=l_c,C=C_c}=0=\frac{\partial^2 T}{\partial r_+^2}\biggr\rvert_{r_+=r_c,l=l_c,C=C_c}\, ,
\end{equation}
which is equivalent to the system
\begin{align}
   \frac{3 k l_c^2 r_c^{-3 \omega -4} \left(Q^2 r_c^{3 \omega }-b r_c \omega  (3 \omega +2)\right)}{64 \pi ^2 C_c}+\frac{3}{4 \pi  l_c^2}-\frac{1}{4 \pi  r_c^2}=0\, \\
   \frac{1}{2 \pi  r_c^3}-\frac{3 k l_c^2 r_c^{-3 \omega -5} \left(4 Q^2 r_c^{3 \omega }-3 b r_c \omega  \left(3 \omega ^2+5 \omega +2\right)\right)}{64 \pi ^2 C_c}=0\, ,
\end{align}
where we also used the definition of the central charge (\ref{centralcharge}) and the relation (\ref{pressure1}) between $l$ and $P$ to find
\begin{equation}\label{g1}
    G=\frac{1}{8\pi}\sqrt{\frac{3k}{2PC}}=\frac{kl^2}{16\pi C}\, .
\end{equation}

These critical outer horizon radius $r_c$, cosmological constant $l_c$, central charge $C_c$ and temperature $T_c=T(r_c,l_c,C_c)$ can only be found for specific values of the Kiselev parameter $\omega$. So, let us analyze the same specific cases of the previous section.

\begin{itemize}
 \item \bm{$\omega=-1/3$}: The critical radius, cosmological constant, temperature and central charge are found to be 
 \begin{equation}
 r_c=\frac{\sqrt{6G}Q}{\sqrt{1-Gb}}\, ,\ \ l_c=\frac{6\sqrt{G}Q}{1-Gb}\, , \ \  T_c=\frac{\sqrt{6}(1-Gb)^{3/2}}{18\pi \sqrt{G}Q}\, , \ \ C_c=\frac{9kQ^2}{4\pi(1-Gb)^2}\, .
 \end{equation}

Notice that due to the coupling with the Kiselev parameter $b$, the critical central charge $C_c$ now depends on $G$, which may indicate the violation of the universality that was presented in \cite{Cong:2021fnf}, since this may be translated into a pressure dependence. In fact, using Eq.(\ref{pressure1}) and $l_c$, we find
\begin{equation}\label{GPQ1}
    G=\frac{1}{b+\sqrt{96\pi P}Q}\, ,
\end{equation}
which can be independently varied for fixed $Q$ and $b$. Besides that, the other critical quantities are the same that are found considering the usual extended phase space thermodynamics \cite{GBKiselev}.

\item \textbf{Average quintessence} \bm{$\omega=-2/3$}: In this case, we have 
\begin{equation}
    r_c=\sqrt{6G}Q\, , \ \ l_c=\sqrt{6}r_c\, , \ \ T_c=\frac{\sqrt{6}}{18\pi\sqrt{G}Q}-\frac{Gb}{2\pi}\, , \ \ C_c=\frac{9kQ^2}{4\pi}\, .
\end{equation}
Notice that $r_c$ and $l_c$ do not couple to $b$, and that $T_c$ only suffers a translation, which also is consistent with previous criticality studies in the standard extended phase space \cite{GBKiselev}. Now, $C_c$ does not depend $G$ and one recovers the universal behavior and also the same relation between $G$ and $P$ of \cite{Cong:2021fnf}, given by (\ref{GPQ1}) if $b=0$.
   
\item \textbf{Average phantom fluid} \bm{$\omega=-4/3$}: 
\begin{align}
    l_c=\frac{\sqrt{2}\sqrt{3Q^2r_c^2+2br_c^7}}{\sqrt{Q^2+4br_c^5}}\, , \ \ T_c=\frac{2Q^2+3br_c^5}{6\pi Q^2r_c+4\pi br_c^6}\, , \ \ C_c=\frac{k(3Q^2+2br_c^5)^2}{4\pi (Q^2+4br_c^5)}\, ,
\end{align}

where $r_c$ satisfies the equation $4Gbr_c^5-r_c^2+6GQ^2=0$ that cannot be solved analytically. In any case, one also loses universality for the critical radius, since variations in $G$ imply in variations in $r_c$. Consequently, this implies in absence of universality of $l_c$, $T_c$ and of $C_c$.
   
\end{itemize}

\begin{figure}
    \centering
    \includegraphics[scale=0.9]{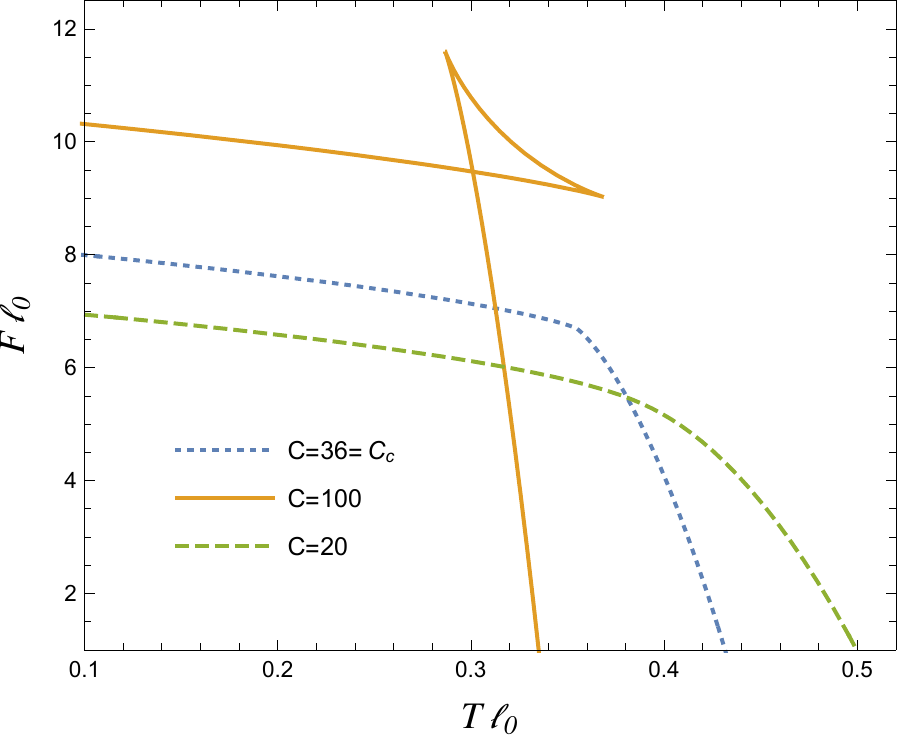}\caption{Depiction of the Free energy versus temperature $F\ell_0\times \, T\ell_0$ for the average quintessence $\omega=-2/3$. We set $Q=1$, $P\ell_0^4=15$, $b\ell_0^3=0.3$ and the central charge parameter $k=16\pi$. The values of the central charge are described in the figure.}\label{fig:FxT}
\end{figure}

We can see these transitions from the behavior of the Free Energy and the Temperature
\begin{align}
    F=M-TS=\frac{3GQ^2l^2+l^2r_+^2-r_+^4}{4Gl^2r_+}-\frac{(3\omega+2)b}{4r_+^3}\nonumber\\
    =\frac{2\pi r_+}{3}\sqrt{\frac{6PC}{k}}-\frac{2\pi Pr_+^3}{3}+\frac{3Q^2}{4r_+}-\frac{(3\omega+2)b}{4r_+^{3\omega}}\, ,\\
    T=\frac{1}{4\pi r_+}+\sqrt{\frac{6kP}{C}}\frac{r_+}{8\pi}-\sqrt{\frac{6k}{CP}}\frac{Q^2}{64\pi^2r_+^3}+\sqrt{\frac{6k}{CP}}\frac{3\omega}{64\pi^2}\frac{b}{r_+^{2+3\omega}}\, ,
\end{align}
where we used equation (\ref{g1}) and (\ref{pressure1}) to express the terms with $G$ and $l$ in terms of variables $P$ and $C$. In Fig.(\ref{fig:FxT}), we depict the free energy as a function of the temperature for the case of an average quintessence fluid $\omega=-2/3$, again in units of a fiducial length in $\ell_0$. As can be seen, the blue/dotted curve describes the critical curve, from which phase transitions can occur in the region above in the phase space $F\times T$, as can be seen in the orange/solid line. Below the critical curve, for smaller values of the central charge, no transitions are allowed (green/dashed curve). From this figure, one realizes the fundamental role played by the central charge in the determination of criticality, which is a novel featured highlighted in Ref.\cite{Cong:2021fnf}. For other values of parameter $\omega$ the free energy follows a similar behavior, therefore, we do not depict them here.

\section{Concluding remarks}\label{sec:conclusion}
Based on the work of Cong, Kubiz\v{n}\'ak and Mann \cite{Cong:2021fnf}, we analyzed a new proposal of extended phase space for the thermodynamics of black holes, which is compatible with the holographic program. It is possible, by the introduction of a novel thermodynamic variable that depends on the gravitational constant (which is allowed to vary), the bulk cosmological constant and the spacetime dimension. In particular, we extended that work for the case of different matter contents that surround a static black hole, modeled by the Kiselev fluid, which is labeled by a free parameter.
\par
Our work demonstrated the validity of the holographic technique for the class of Kiselev black holes. We generalized the bulk first law of thermodynamics by the introduction of an alternative thermodynamic volume and a chemical potential-like term $\mu$, which is conjugate to the central charge. As an application, we analyzed the $4$-dimensional case assuming three different average equations of state parameters ($\omega=-1/3\, , -2/3\, , -4/3$). For which we studied their relationships between the sign of the chemical potential and the size of the black hole. Thus, when the reverse isoperimetric inequality is valid for the standard volume $V$, then the black holes follow a simple size classification based on the sign of the chemical potential: positive (negative) $\mu$ for small (large) black holes. On the other hand, if this inequality is no longer valid for $V$, then such classification is broken since the polynomial that regulates the sign of $\mu$ has a more diverse behavior. 
\par
We also analyzed phase transitions, in which the central charge plays a fundamental role in the determination of the criticality, thus in agreement with results of \cite{Cong:2021fnf}. However, we verified an important divergence between ours and those previous results:  a possible surrounding matter for which the universality of the critical central charge is no longer valid.  This property is violated as a result of the dependence of the critical central charge on the gravitational constant due to the coupling with Kiselev's fluid. We also illustrated the behavior of the phase transitions from the phase space diagram of the free energy versus the temperature. We observed that the critical central charge plays the role of separating the space between regions in which phase transition may or may not occur.


\section*{Acknowledgements}
I. P. L. would like to acknowledge the contribution of the COST Action CA18108. I. P. L. was partially supported by the National Council for Scientific and Technological Development - CNPq grant 306414/2020-1. R. B. A. was partially supported by the Tutoring Program of the Federal  University  of  Lavras.

\section*{Data Availability Statement}
Data sharing is not applicable to this article as no new data were created or analyzed in this study.


\bibliographystyle{utphys}
\bibliography{variableG}
\end{document}